\documentclass{aa}
\usepackage{graphicx,epsfig}
\usepackage{subfigure}
\usepackage{natbib}
\usepackage[usenames]{color}
\usepackage{txfonts}
\usepackage{url}
\usepackage[dvips]{hyperref}
\usepackage{stfloats}

\newcommand{\FeIX}{\ion{Fe}{ix}}
\newcommand{\FeXII}{\ion{Fe}{xii}}
\newcommand{\FeXIV}{\ion{Fe}{xiv}}
\newcommand{\kms}{km s$^{-1}$}
\newcommand{\aia}{{AIA/SDO}}

\begin{document}

\title{Propagating intensity disturbances in polar corona as seen from AIA/SDO}
\author{S. Krishna Prasad\inst{1}, D. Banerjee\inst{1}, G. R. Gupta\inst{1,2}}
\institute{Indian Institute of Astrophysics, Koramangala, Bangalore 560034, India. \and Joint Astronomy Programme, Indian Institute of Science, Bangalore, 560012, India.}
\offprints{S. Krishna Prasad \email{krishna@iiap.res.in}}
\date{Received ... / Accepted ...}
\abstract
{Polar corona is often explored to find the energy source for the acceleration of the fast solar wind. Earlier observations show omni-presence of quasi-periodic disturbances, traveling outward, which is believed to be caused by the ubiquitous presence of outward propagating waves. These waves, mostly of compressional type, might provide the additional momentum and heat required for the fast solar wind acceleration. It has been conjectured that these disturbances are not due to waves but high speed plasma outflows, which are difficult to distinguish using the current available techniques.}
{With the unprecedented high spatial and temporal resolution of \aia, we search for these quasi-periodic disturbances in both plume and interplume regions of the polar corona. We investigate their nature of propagation and search for a plausible interpretation. We also aim to study their multi-thermal nature by using three different coronal passbands of AIA.}
{We chose several clean plume and interplume structures and studied the time evolution of specific channels by making artificial slits along them. Taking the average across the slits, space-time maps are constructed and then filtration techniques are applied to amplify the low-amplitude oscillations. To suppress the effect of fainter jets, we chose wider slits than usual.}
{In almost all the locations chosen, in both plume and interplume regions we find the presence of propagating quasi-periodic disturbances, of periodicities ranging from 10-30 min. These are clearly seen in two channels and in a few cases out to very large distances ($\approx 250$\arcsec) off-limb, almost to the edge of the AIA field of view. The propagation speeds are in the range of 100-170~\kms. The average speeds are different for different passbands and higher in interplume regions.}
{Propagating disturbances are observed, even after removing the effects of jets and are insensitive to changes in slit width. This indicates that a coherent mechanism is involved. In addition, the observed propagation speed varies between the different passpands, implying that these quasi-periodic intensity disturbances are possibly due to magneto-acoustic waves. The propagation speeds in interplume region are higher than in the plume region.}


\titlerunning{Propagating disturbances in polar corona}
\authorrunning{S. Krishna Prasad et~al.}
\maketitle

\section{Introduction}
During solar minimum, polar corona is structured with several bundles of open magnetic field regions called plumes. The dark regions between plumes are called interplumes. Plumes and interplumes are characterized by slightly different properties. Plumes are denser and cooler than interplume regions \citep[e.g.,][]{2006A&A...455..697W}. Interplume regions exhibit broader line profiles than plumes \citep{2000SoPh..194...43B}. \citet{1997ApJ...491L.111O} first reported the quasi-periodic variations in polar plumes using the white light channel of UVCS/SoHO. They proposed that they are signatures of compressional waves. \citet{1998ApJ...501L.217D} identified the ubiquitous presence of these propagating disturbances using EIT/SoHO. \citet{1999ApJ...514..441O,2000ApJ...533.1071O} interpreted these disturbances as slow magneto-acoustic waves. Similar work has been performed by many others using spectroscopic observations of CDS/SoHO \citep{2000SoPh..196...63B,2001A&A...377..691B,2001A&A...380L..39B,2006A&A...452.1059O, 2007A&A...463..713O} and more recently using SUMER and EIS \citep{2009A&A...499L..29B, 2010ApJ...718...11G}, and all concluded that these disturbances were magneto-acoustic waves. From STEREO observations, \citet{2010A&A...510L...2M} suggested that these are quasi-periodic high speed outflows rather than waves. A clear understanding of the basic physical mechanism involved is essential to quantify the significance and contribution of these quasi-periodic disturbances to the acceleration of the fast solar wind and/or coronal heating. It has also been debated whether the plumes \citep{1999JGR...104.9947C,2003ApJ...589..623G,2005ApJ...635L.185G} or interplumes \citep{2000A&A...353..749W,2003ApJ...588..566T,2010ApJ...718...11G} are the preferred channels for the acceleration of the fast solar wind. In this letter, we use the data from \aia~ in three coronal channels, to explore and study propagating quasi-periodic disturbances using space-time maps, constructed from several plume and interplume regions and try to address these problems. The simultaneous full-disk imaging capability of \aia~ in different channels with high spatial and temporal resolution, made this study possible. In the following sections, we first discuss the observational data used in this study, then the techniques used for the analysis and results obtained, and finally draw our conclusions.
\section{Observations}
We use the imaging data from the Atmospheric Imaging Assembly on-board SDO \citep[AIA;][]{2010soph...sub} in the Extreme Ultraviolet regime in three coronal channels centered at wavelengths 171~\AA, 193~\AA, and 211~\AA. These three coronal channels correspond to their emissions mainly from \FeIX, \FeXII, and \FeXIV~ and their temperature responses peak at 0.8 MK, 1.25 MK, and 1.6 MK, respectively. Hereafter, we use the central wavelengths to refer to these channels. Data used in current study is spread over two hours on July 20, 2010 from 02:00 UT to 03:59 UT. It is Rice compressed, calibrated level 1.0 data and is directly used in this analysis without applying any further corrections. We did not perform any de-rotation since the rotation effect would be insignificant on our analysis. The AIA pixel size is $\approx$0.6\arcsec~and the cadence is 12 s. Fewer than ten frames out of 600 frames from these two hours, in each passband are missing, which are added by linear interpolation. We chose to study the south polar region because it contains clear plume structures. 
\section{Analysis \& results}
\begin{figure*}
\sidecaption
  \includegraphics[width=12cm]{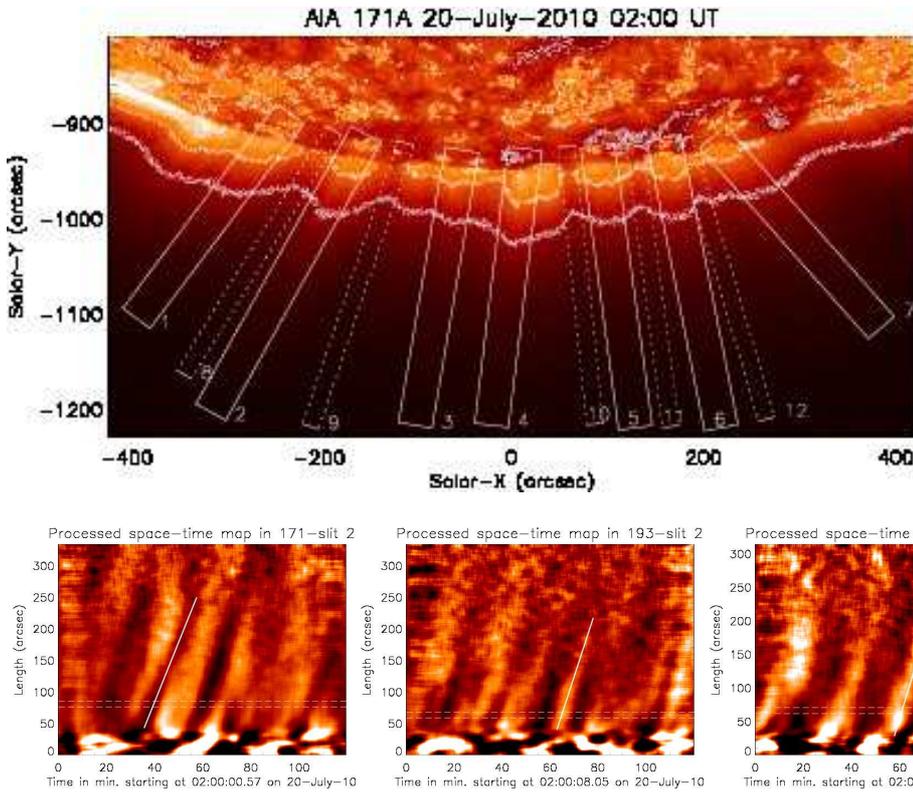}
\caption{South polar region of the Sun as seen by \aia~ through one of its EUV channels centered at 171~\AA. This snapshot is taken at 02:00 UT on 20 July 2010. Overplotted dotted curves are the contours of three different intensity levels indicating clear plume and interplume structures. The rectangular boxes delineate the locations of artificial slits extracted for the analysis, following intensity contours. Slits over plume regions are 60 pixels ($\approx$36\arcsec) wide and those over interplume regions are 30 pixels ($\approx$18\arcsec) wide. These are marked differently using solid lines for plume locations and dotted lines for interplume locations. Slit numbers are labeled at the bottom right of each slit. The movies show the temporal evolution as seen in the 171~\AA, 193~\AA, and 211~\AA~channels.}
  \label{f1}
 \end{figure*}
\begin{figure*}[ht]
\centering
\includegraphics[angle=90,width=4.5cm]{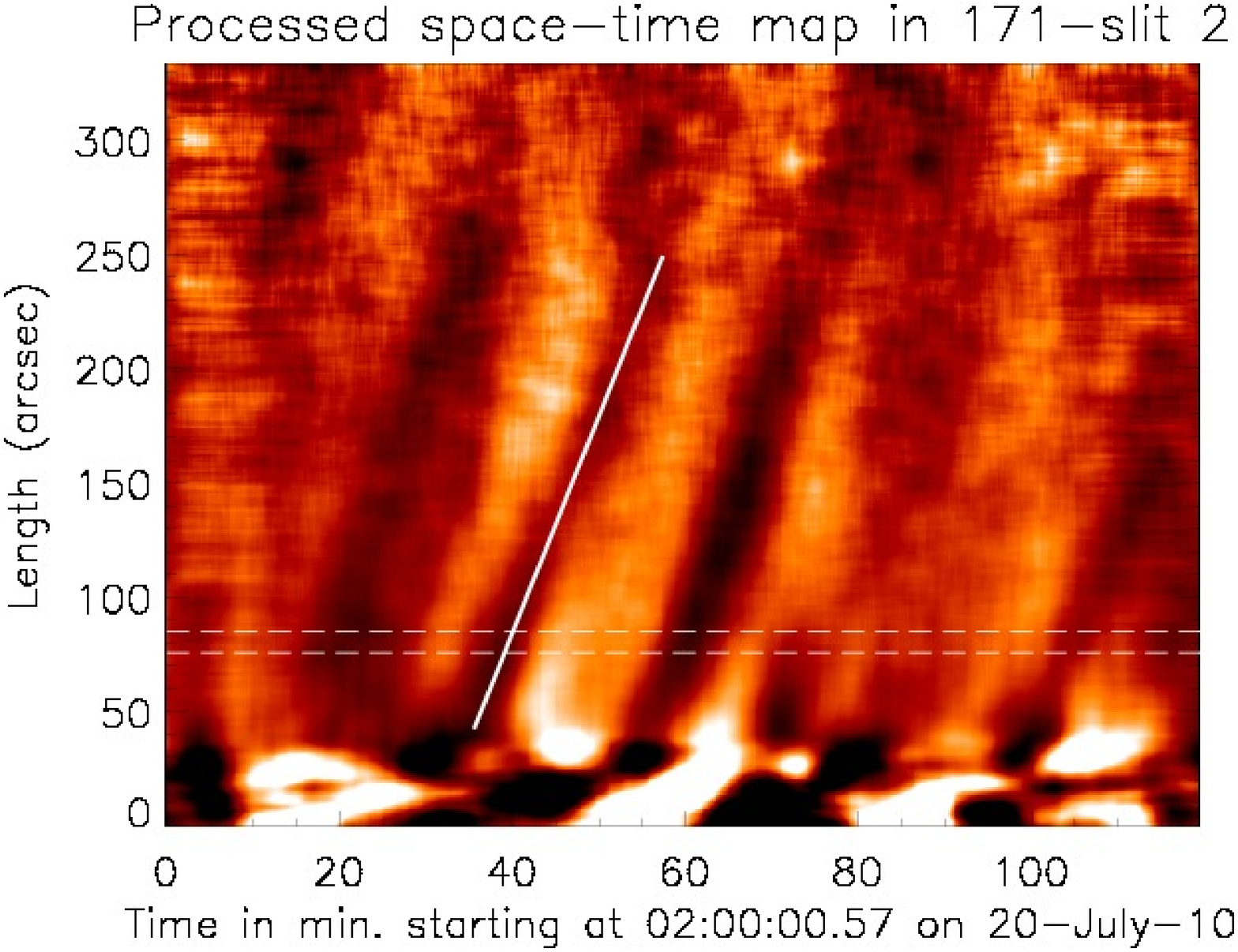}
\includegraphics[angle=90,width=4.5cm]{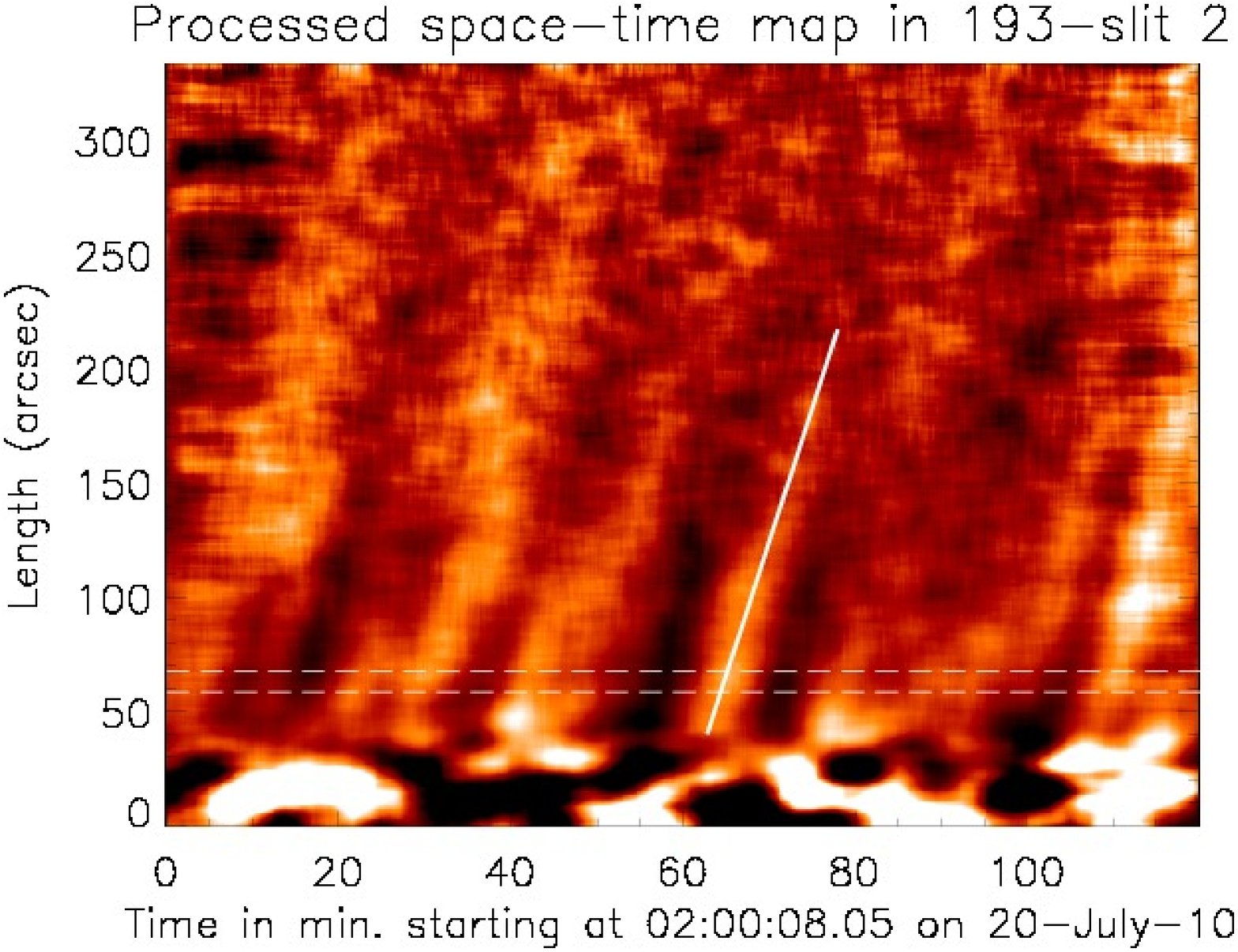}
 \includegraphics[angle=90,width=4.5cm]{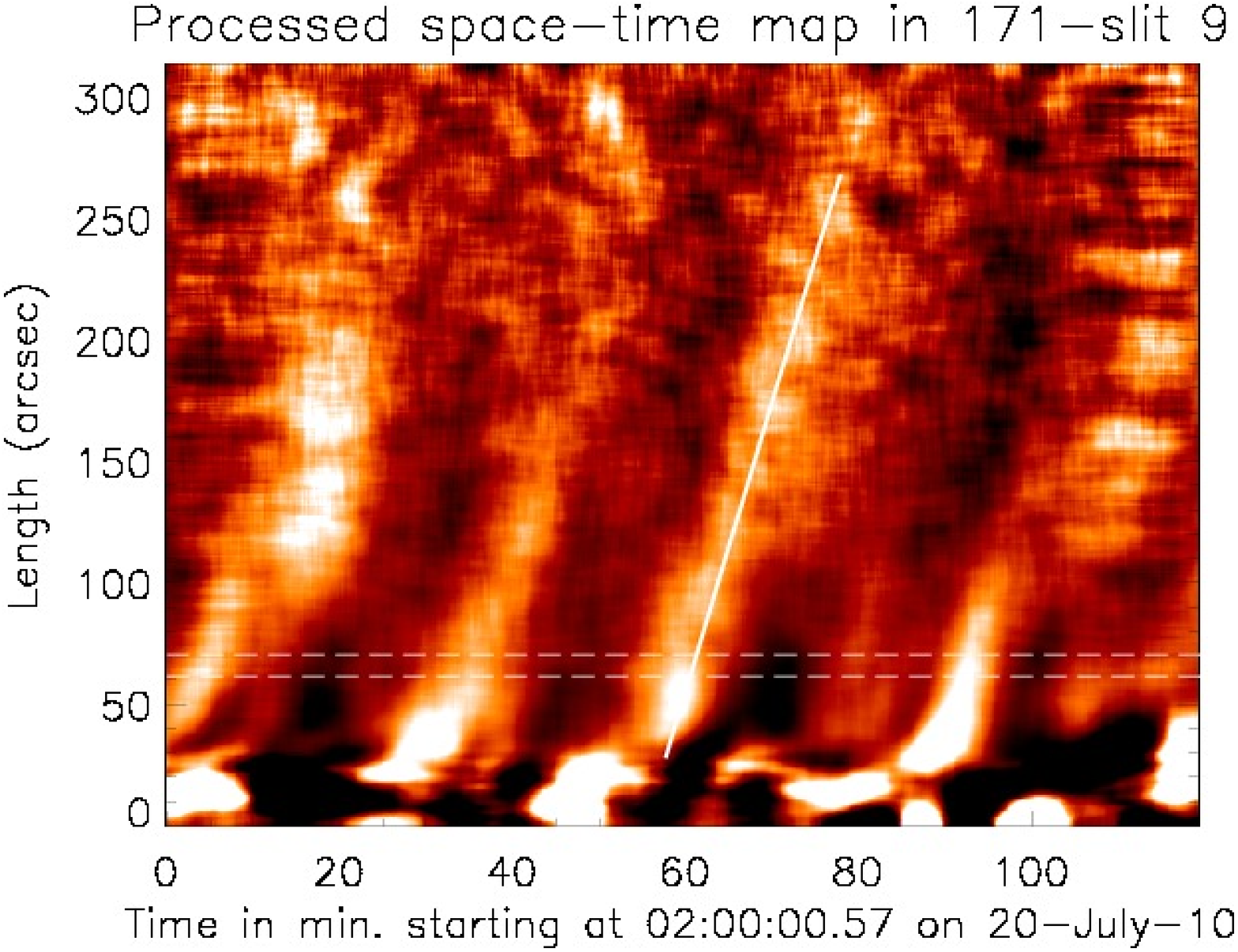}
 \includegraphics[angle=90,width=4.5cm]{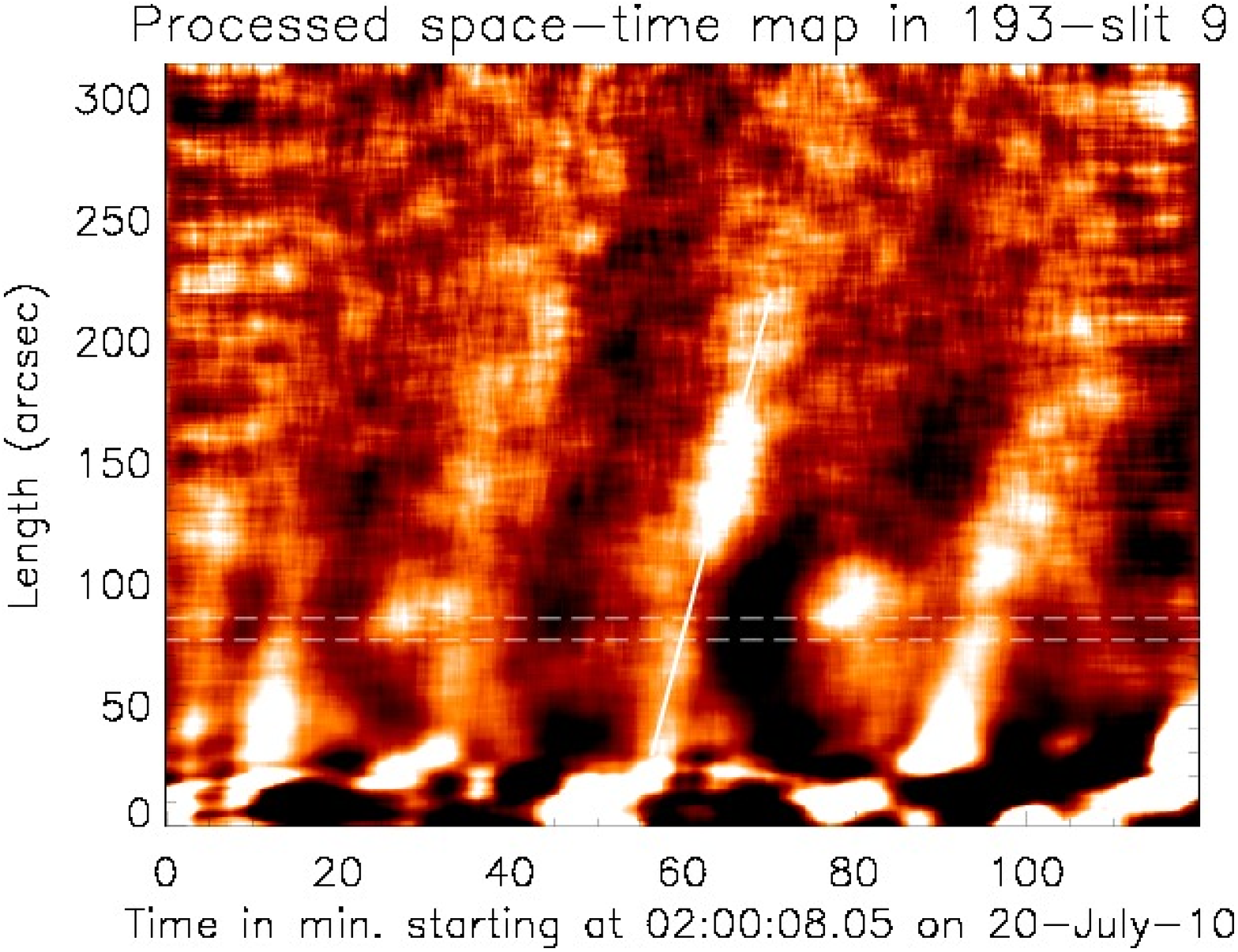}
\caption{Space-time maps with time on X-axis, constructed from slits 2 and 9 (See Fig.~\ref{f1}) and processed, in two coronal channels of AIA, 171~\AA~ and 193~\AA. First two panels are for slit 2 (plume) and the last two are for slit 9 (interplume). The slanted solid line, marked in each panel, following the ridges is used for the propagation speed estimation. The horizontal dotted lines in each of these, enclose the rows averaged for wavelet analysis. Corresponding maps in 211~\AA~ are shown in Fig.~\ref{fig4} (available online only).}
\label{f2}
\end{figure*}
\begin{figure*}[ht]
\centering
  \includegraphics[angle=90,width=8.5cm,clip=true]{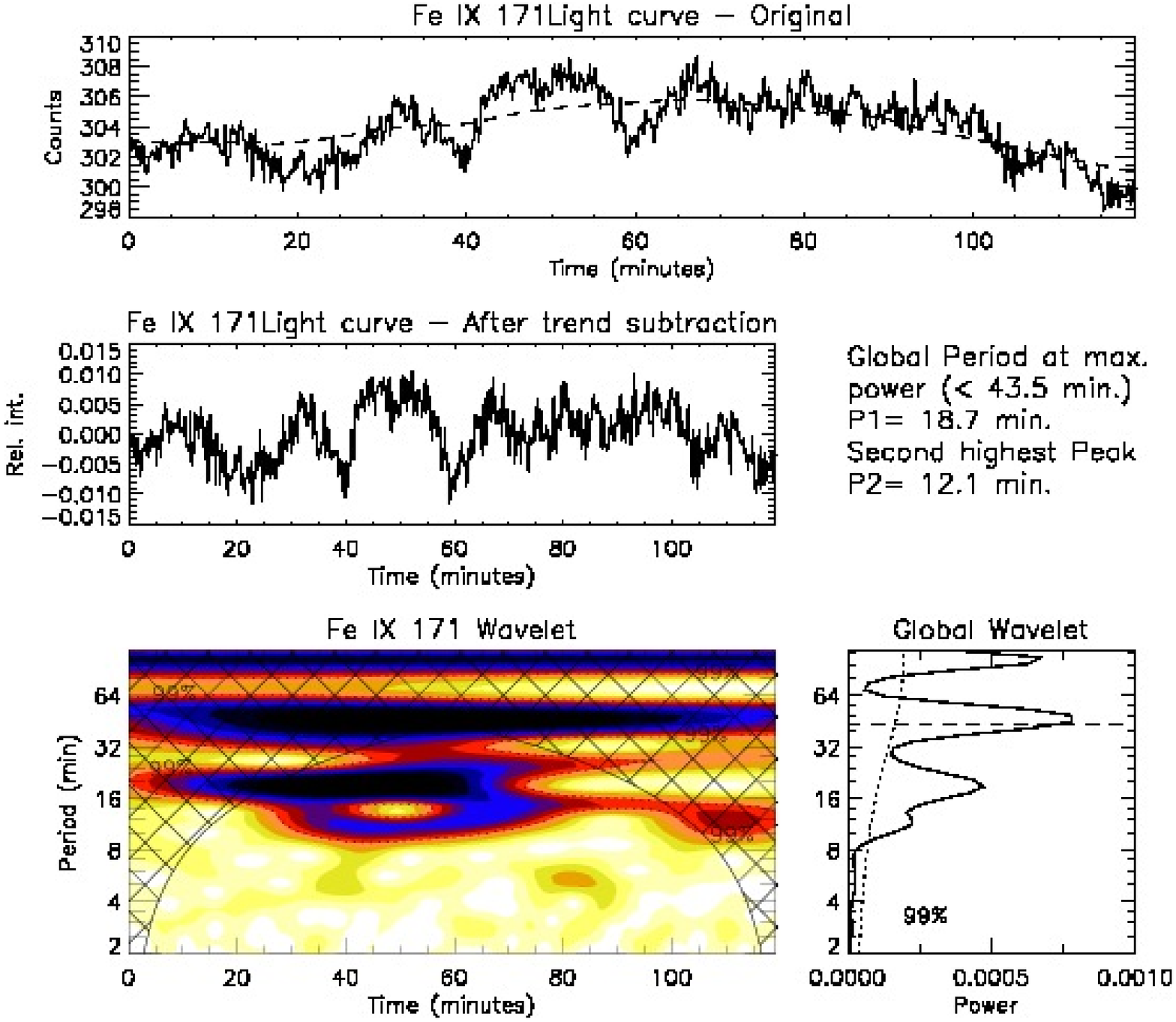}
  \includegraphics[angle=90,width=8.5cm,clip=true]{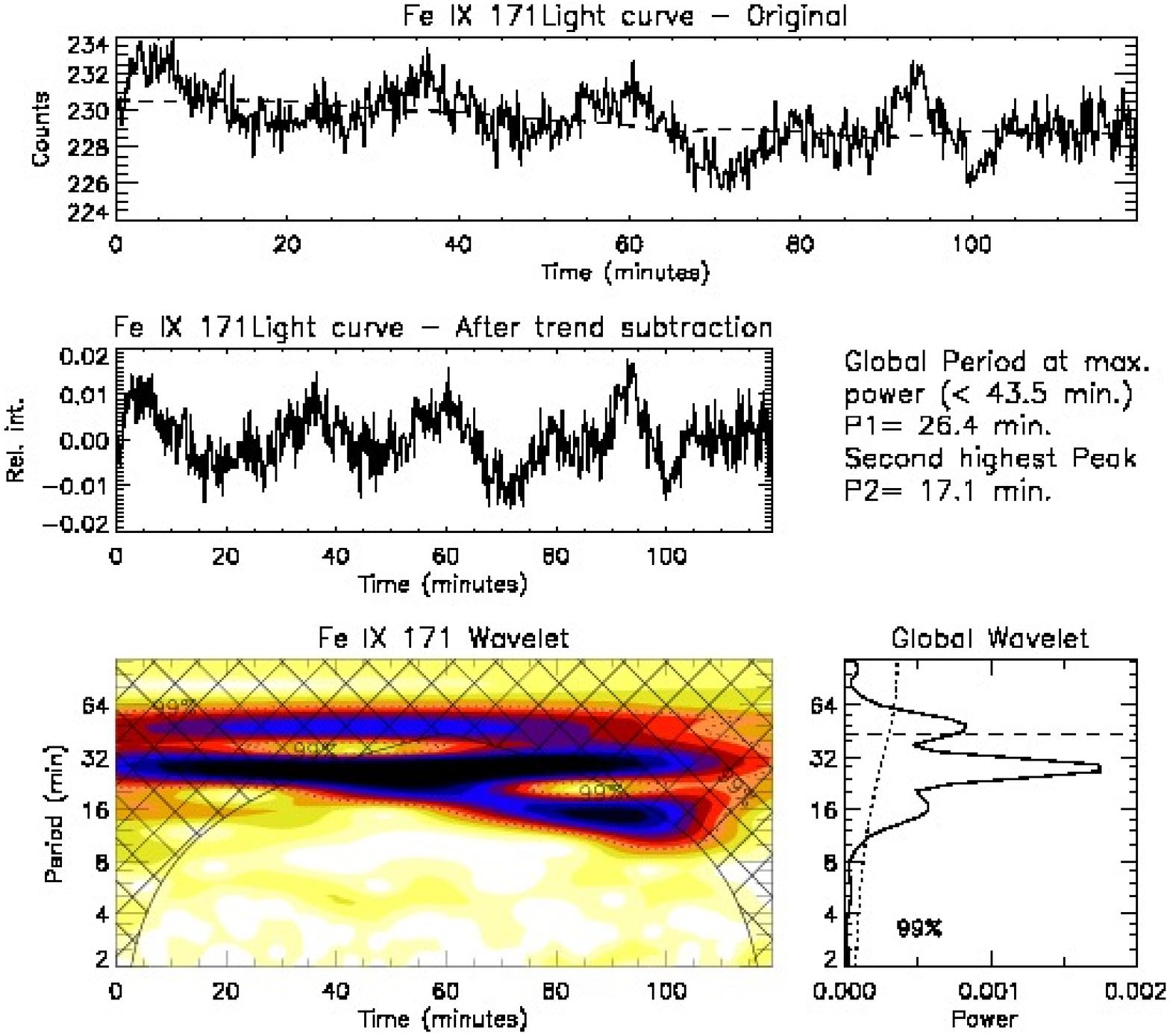}
\caption{Wavelet analysis results for slits 2 and 9 (see Fig.~\ref{f1}) in 171~\AA. Top two panels in each of these show the light curves, original and background-trend-subtracted, for the region enclosed by horizontal lines in Fig.~\ref{f2}. The bottom left panel shows the wavelet plot with contours enclosing the 99 \% confidence regions, for a white-noise process. Bottom right panel is the global wavelet plot with 99 \% global confidence level, overplotted as a dotted line. The periods of primary and secondary peaks are also written in the text (middle-right). Corresponding plots in 193~\AA~ are shown in Fig.~\ref{fig5} (available online only)}
\label{f3}
\end{figure*}

In the south polar region, several structures are identified following peaks (plumes) and depressions (interplume) in intensity using contours (see Fig.~\ref{f1}). At these locations, artificial slits of 60  pixels ($\approx$36\arcsec) width in plume region and 30 pixels ($\approx$18\arcsec) width in interplume region are extracted from each time frame, averaged over width and stacked to construct space-time maps. We chose a total of 12 slits, seven in plume and five in interplume regions, covering the on-disk part of the plumes wherever visible and extended almost to the end of the AIA field of view. The intensity contours and chosen slits with their exact widths are overplotted with solid (plume) and dotted (interplume) lines on a snapshot in 171~\AA~ in Fig.~\ref{f1}. The slit numbers are marked at the bottom right of each slit. Different widths are chosen to isolate interplume regions from the plume regions, hence study them separately. Same locations are used in all the three passbands as shifts of a few pixels between them (because of any misalignment), if at all present, can be ignored. The space-time maps thus obtained for each slit are then processed by smoothing over time using 200 points ($\approx$40 min) smoothing, subtracting from the original, and normalizing to the smoothed image. The smooth subtraction enhances the faint variations and the normalization will take care of the radial intensity fall. Any quasi-periodic disturbance in intensity, propagating along the slit will appear as alternate bright and dark ridges in this processed map. These processed maps for slit 2 (plume) and slit 9 (interplume), can be seen in Fig.~\ref{f2}. One can see from Fig.~\ref{f2} first panel, that one such disturbance traveled almost 250\arcsec off-limb, reaching the end of slit, and might have traveled yet further.
The slope of these slanted ridges indicates the projected propagation speed of the disturbance, and the spacing  between two successive ridges implies the periodicity. Since these ridges are fainter and rarely seen in 211~\AA, we limit our analysis to 171~\AA~ and 193~\AA~. The speeds are calculated by fitting straight lines to the ridges and periodicity calculations are performed using a wavelet analysis. The horizontal lines marked on Fig.~\ref{f2} enclose the region averaged and used in our wavelet analysis. We use the Morlet function as the mother wavelet \citep[see][]{1998BAMS...79...61T}. The results of the wavelet analysis for slits 2 and 9 can be seen in Fig.~\ref{f3}. 
The top panels of each subfigure in Fig.~\ref{f3} shows the original light curve which sometimes contains longer periods that may be due to background variations. These are removed by choosing the appropriate background from the running average (over 50-60 min) of the original, which are overplotted. The Middle panel shows this background trend subtracted light curve. The description of other panels can be found in corresponding figures. The primary and secondary peaks in periodicities, represented as P1 and P2, are well above the 99~\% confidence level \citep{1998BAMS...79...61T} for a white noise process. The periods and propagation speeds for all the slits are tabulated in Table~\ref{table1}. If the second peaks are not available or below the 99~\% significance level, they are marked as '\dots'.
\begin{table}
  \caption{Observed periods and projected propagation speeds for all the slits marked in Fig.~\ref{f1}, in both 171~\AA~ and 193~\AA.}
  \label{table1}
  \centering
  \begin{tabular}{c l l l l}
  \hline\hline
            & \multicolumn{2}{c}{\FeIX~ 171~\AA} & \multicolumn{2}{c}{\FeXII~ 193~\AA} \\
  \cline{2-5}
  Slit No.  & P1(P2)      & Speed & P1(P2)      & Speed \\
            & (min)      & (\kms)& (min)      &(\kms) \\
  \hline
       1    & 24.3 (13.2) & 103.2$\pm$1.4 & 28.8 (12.1) & 133.1$\pm$2.3 \\
       2    & 18.7 (12.1) & 113.8$\pm$1.4 & 20.4 (17.1) & 140.2$\pm$2.3 \\
       3    & 31.4 (13.2) & 110.9$\pm$1.5 & 26.4 (13.2) & 124.2$\pm$1.6\tablefootmark{a} \\
       4    & 18.7 (11.1) & 117.5$\pm$1.9 & 22.2 (10.2) & 142.2$\pm$3.2\tablefootmark{b}\\
       5    & 28.8 (14.4) & 109.9$\pm$1.6 & 26.4 (10.2) & 137.6$\pm$2.5 \\
       6    & 26.4 (12.1) & 116.5$\pm$1.7 & 24.3 (17.1) & 132.6$\pm$1.9 \\
       7    & 28.8 (18.7) & 121.5$\pm$2.1 & 31.4 (17.1) & 136.9$\pm$3.4 \\
 \hline
   $\mu\pm\sigma$ &   &113.4$\pm$5.6 & &135.2$\pm$5.6 \\
      $\pm\delta$ &   &     $\pm$0.6 & &     $\pm$0.9 \\
 \hline
       8    & 24.3 (14.4) & 137.4$\pm$2.3 & 24.3 (14.4) & 149.8$\pm$2.6\tablefootmark{a}\\
       9    & 26.4 (17.1) & 142.5$\pm$1.8\tablefootmark{a} & 26.4 (18.7) & 164.7$\pm$2.9\tablefootmark{a}\\
      10    & 18.7 (12.1) & 144.9$\pm$2.7\tablefootmark{b} & 18.7 (11.1) & 153.9$\pm$3.6\tablefootmark{b}\\
      11    & 24.3 (...)  & 127.2$\pm$1.6\tablefootmark{a} & 22.2 (...)  & 167.4$\pm$3.8\tablefootmark{a}\\
      12    & 20.4 (13.2) & 121.2$\pm$2.1 & 28.8 (15.7) & 165.1$\pm$3.9 \\
  \hline
   $\mu\pm\sigma$ &   &134.6$\pm$9.1& & 160.2$\pm$7.0 \\
      $\pm\delta$ &   & $\pm$0.9     & &$\pm$1.5 \\
  \hline
  \end{tabular}
\tablefoot{P1 and P2 are the primary and secondary periods (see Fig.~\ref{f3}). $\mu$ is the average of all the values above, $\sigma$ is the standard deviation representing statistical error, and $\delta$ is the instrumental uncertainty propagated from individual values.
  \tablefoottext{a}{Jet-like features that can be seen from the movies.}
  \tablefoottext{b}{Enhanced intensity regions showing a decelerating ridge at the end of space-time map. This is due to some eruption from a region above slits 4 and 10.}
}
\end{table}
The periodicities are in the range of 10-30 min and the  propagation speeds are 100-170 \kms. The errors in the calculation of speed, are due to instrumental uncertainties, given by $\delta v= v(\frac{\delta x}{x}+\frac{\delta t}{t})$, where $v, x,$ and $t$ are the propagation  speed, and the vertical and horizontal extents of the lines marked in space-time maps, and in addition, $\delta x$, and $\delta t$ are the pixel scale (0.6$\arcsec$) and time cadence (0.2 min), respectively. The errors are small, owing to the high spatial and temporal resolutions of the instrument, but the true uncertainty in the slope measurement might be larger than this. We may alternatively consider the width of the strip, in time direction as $\delta t$, but despite choosing the narrower ones, the uncertainties are too large and the actual human error would be definitely smaller than this. Hence, we consider the standard deviation ($\sigma$) over the average propagation speed ($\mu$) of all the values as the measurement error. Considering this, there is a noticeable difference in the average speeds between plume and interplume regions and also between different passbands. \\

Mostly in the 193~\AA\ passband, we detected in a few cases, one or two ridges that are steeper and brighter than other ridges. In one instance, we also observed a decelerating bright ridge at the end of the time sequence, which is brightest in 211~\AA. The former case is identified as the effect of strong jets and the latter is due to some eruption that occurred just above the slits 4 and 10 and propagated off-limb. The slits affected by these events are marked in Table~\ref{table1}. The eruption can be seen most clearly from the processed movies available online\footnote{\url{ftp://ftp.iiap.res.in/krishna/aia_polar}}. Strong jets should be identifiable in the {\em original} movies. We emphasize {\em original} here, since the movies, processed to extract the fainter variations ($< 5 \%$) may produce the same visual impression for both waves and jet outflows. We note that the fainter jets, if at all present, do not affect our analysis because of our wider slits. 
\section{Discussion}
We discuss various possible explanations of our results and a plausible interpretation. We chose wider slits  than usual to avoid the effects of fainter (few \% above background) jets, if any were present at all. It is appropriate here to note that, though we present the case of 60 (plume) and 30 (interplume) pixel widths, we found the same results for slits as wide as 90 pixels ($\approx$54\arcsec) and as narrow as 15 pixels ($\approx$9\arcsec) which implies that a coherent mechanism is involved. Since the jets cannot be produced so coherently, the wider slits will average out the effect of fainter jets, but the strong jets may remain. Using simple calculations, we roughly estimate that a jet of a five pixel ($\approx$3\arcsec) width and of intensity 10 \% above background will become a $< 1 \%$ variation assuming there is one jet at a time inside the 60 pixel slit. Hence, the stronger ($> 10 \%$) jets may appear in our space-time maps, but can be easily identified from the movies and filtered out. We found few such examples, most of which were easily identified from the presence of distinct ridges with steeper slopes that are indicative of higher speeds. We also came across a faster jet that was indistinguishable from the other ridges because of its inclination to our slit 9. These cases increase the ambiguity and shall be avoided by a careful inspection of movies. Following these steps carefully, we avoid the case of jets, yet still find propagating quasi-periodic disturbances in almost all cases that more importantly are insensitive to changes in slit width within certain limits. \citet{1997SoPh..175..393D} suggested that the ionization temperature of the plume and interplume regions is around 1.0-1.5 MK. The propagating disturbances we find can also be seen clearly in 171~\AA~ and 193~\AA~ but rarely in 211~\AA, whereas the jets are seen mostly in 193~\AA~ and 211~\AA~. This might be because jets are streams of hotter (than background) material whereas the (compressive) waves are local modulations in density. Another point is, though the periodicity values range from 10 min to 30 min there being three periodicities in most cases, with the first one varying between $\approx$12 min and $\approx$18 min, the second one in-between 18 min, 24 min, and 30 min, and the third one around 45 min (though, inside the cone of influence, see global wavelet plots in Fig.~\ref{f3}). The repeated presence of these selective periodicities might be indicative of harmonics. Furthermore, the difference in the propagation speeds between 171~\AA~ and 193~\AA~ cannot be explained by a multi-thermal jet or outflow scenario whereas the temperature dependence of acoustic speed can explain the difference. The acoustic speeds are related to temperature by the formula, $C_{S}\approx152~T^{1/2}~m~s^{-1}$ with $T$ in K, as given by \citet{1984smh..book.....P}. The ratio of observed speeds is 1.19, which can be compared to the acoustic speed ratio 1.25 considering the peak temperatures of 171~\AA~ (0.8~MK) and 193~\AA~(1.25~MK). Now, comparing theoretically estimated acoustic speeds in 171~\AA~(136~\kms) and 193~\AA~(170~\kms) to those observed average values listed in Table~\ref{table1}, it is suggested that these disturbances are of slow magneto-acoustic type. However, if the actual uncertainties in speed calculations are larger (about 20\%) then the difference in speed will be negligible and the interpretation in terms of acoustic type may not be valid. We also attempted to use the 211~\AA~ and 171~\AA~ pair, since their acoustic speed ratio is large. But from the space time maps corresponding to 211~\AA~ line, the ridges are clearly visible only for slit 2 location (see Fig~\ref{fig4}). The measured propagation speed for slit 2 in 211~\AA~is 142.3$\pm$4.4~\kms\, and the speed ratio with respect to 171~\AA~ is 1.25 compared to the theoretical value 1.41. For the plume and interplume comparison, the propagation speeds are slightly higher in interplumes but we found no noticeable difference in the ridge pattern. In some interplume cases, though there are slight curvatures at the bottom, close to the limb, they could be more due to projection effects than acceleration unlike the case reported by \citet{2010ApJ...718...11G}.

\section{Conclusion}
We have traced several plume and interplume structures using contrast-enhanced images recorded in three coronal channels 171~\AA, 193~\AA, and 211\AA~ of \aia~. The excellent spatial and temporal resolution of \aia, using space-time maps, has enabled us to find ubiquitous presence of quasi-periodic disturbances, which are  unaltered by any changes in the slit width within certain limits. This implies that these are coherent phenomena. We have suppressed fainter jets using wide slits and excluded the effects of stronger jets by analyzing the movies. We emphasize that although our observations imply that these disturbances can be mostly due to presence of waves, the flow scenario cannot be completely ruled out based on these imaging observations. To explore the exact nature of these propagating disturbances, one needs to use coordinated imaging and spectroscopic observations, as demonstrated by \citet{2010ApJ...722.1013D} and \citet{2011ApJ...727L..37T}. We find that these disturbances travel far off-limb ($\approx$250\arcsec) and may provide additional momentum to aid the acceleration of fast solar wind in coronal holes. We do not find any evidence of either acceleration or deceleration in either plume or interplume regions, but the propagation speeds are higher and close to the acoustic speed in interplume regions.

\begin{acknowledgements}
We would like to thank the anonymous referee for valuable comments which enabled us to improve the quality of the paper. The AIA data used here is the courtesy of SDO (NASA) and AIA consortium. We thank David Boyes and Veronique Delouille of Royal Observatory of Belgium, for their help in accessing the data. 
\end{acknowledgements}

\bibliographystyle{aa}
\bibliography{16405ref}
%

\Online
\onlfig{4}{
\begin{figure*}[h] 
\centering 
\includegraphics[angle=90,width=8.5cm]{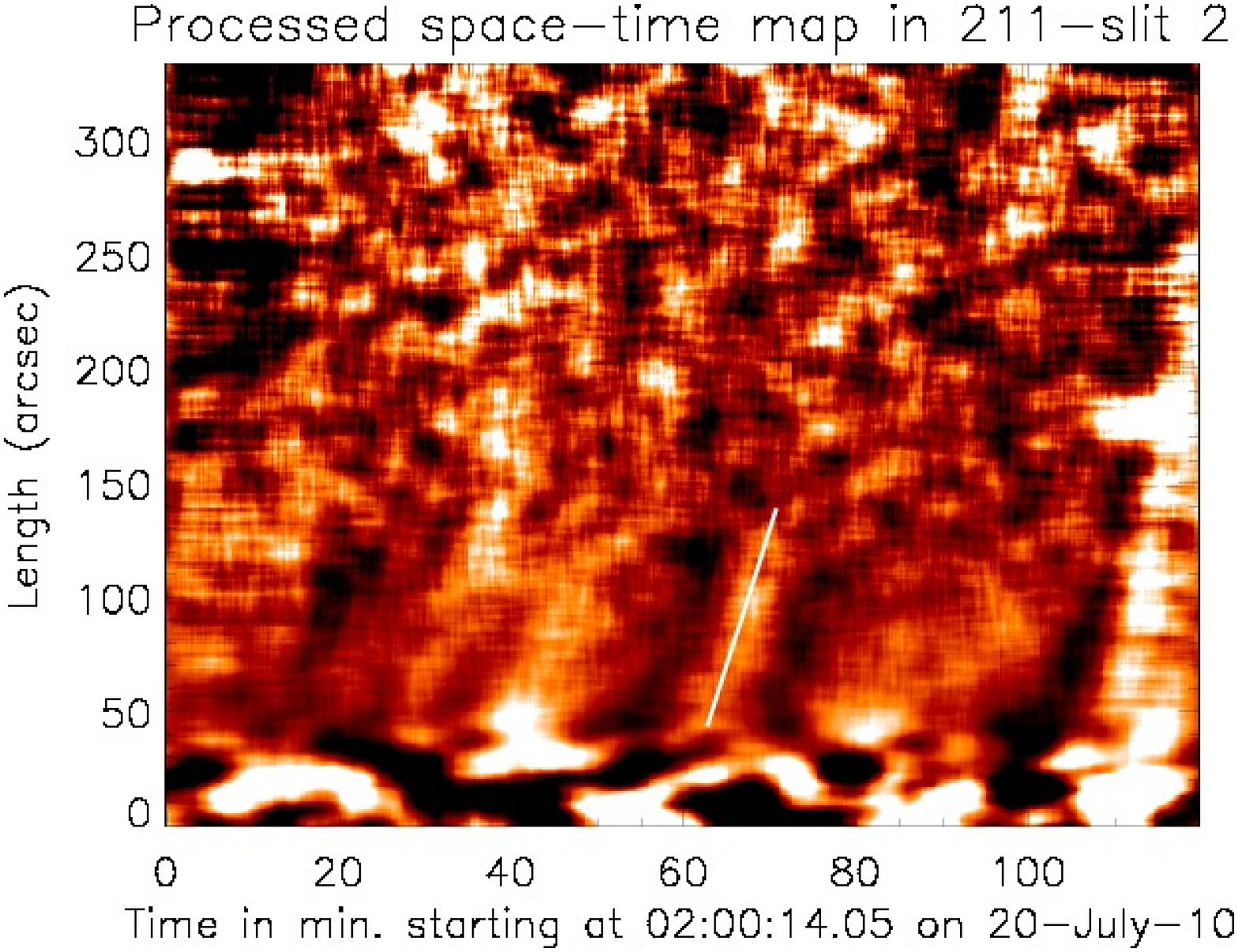}
\includegraphics[angle=90, width=8.5cm]{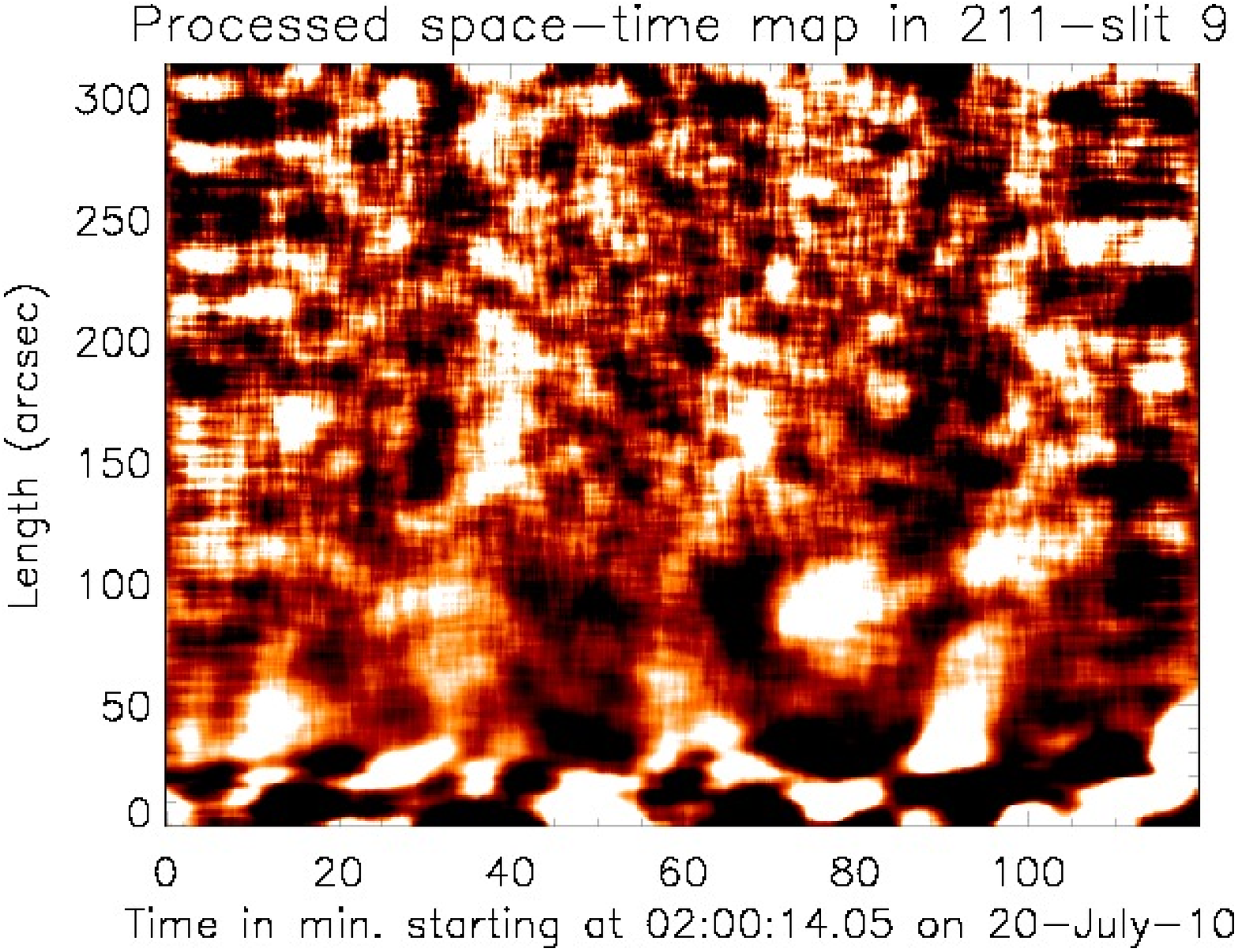}
 \caption{Processed space-time maps in 211~\AA, for slits 2 (left) and 9 (right). The ridges are faintly seen in slit 2 up to $\approx$90\arcsec~off-limb. The slanted solid line overplotted on the left map, following the ridge, is used for propagation speed calculation.}
 \label{fig4}
\end{figure*}
}
\onlfig{5}{
\begin{figure*}[h] 
\centering 
\includegraphics[angle=90,width=8.5cm]{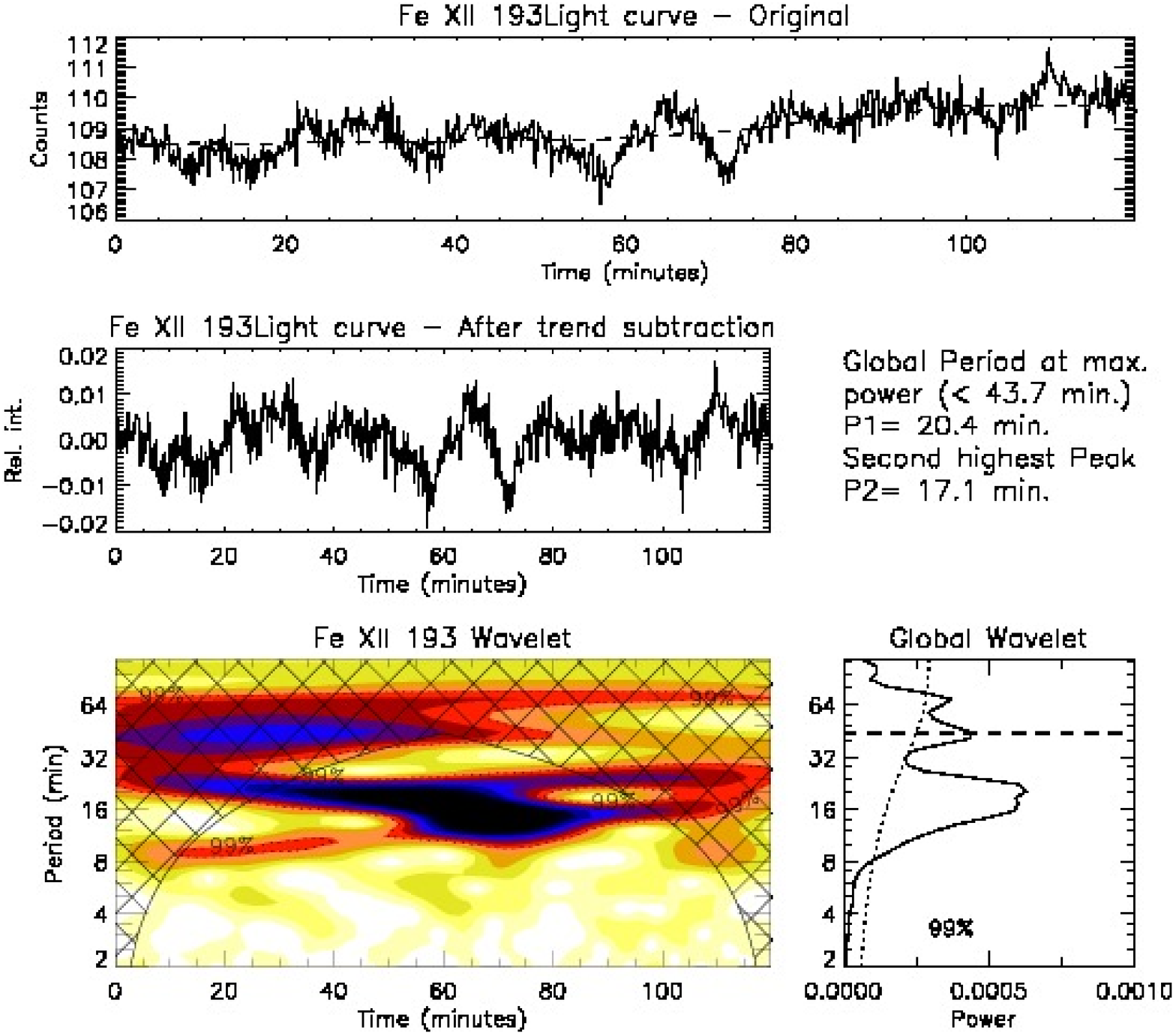}
\includegraphics[angle=90, width=8.5cm]{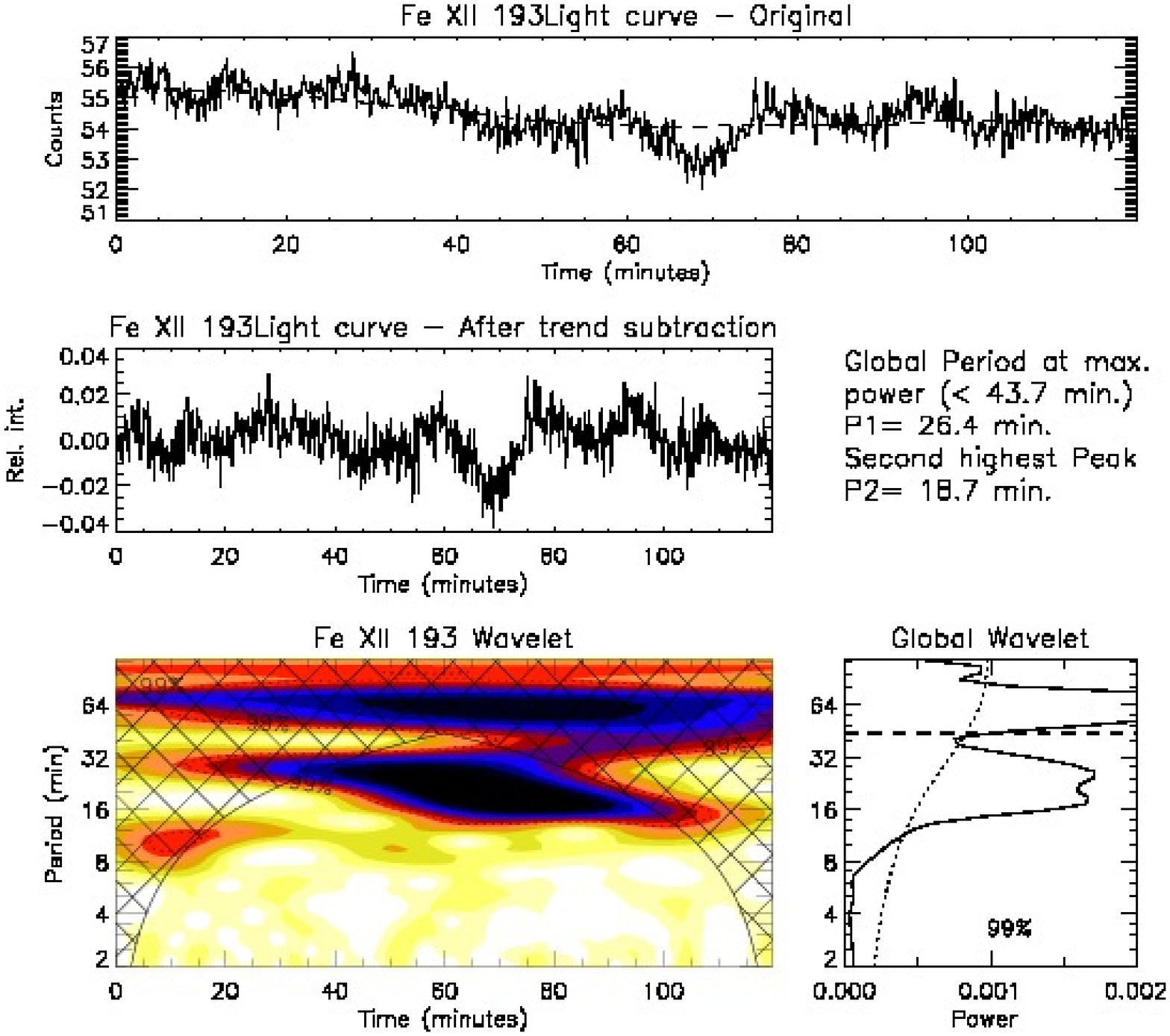}
 \caption{Wavelet plots for slits 2 (left) and 9 (right) in 193~\AA. The description of individual panels is given in Fig.~\ref{f3}}
 \label{fig5}
\end{figure*}
}
\end{document}